\documentclass[letterpaper,twocolumn,10pt]{article}
\usepackage{usenix}

\usepackage{tikz}
\usepackage{amsmath}
\usepackage{mathtools}
\usepackage{amssymb}

\usepackage{algorithm2e}
\usepackage{algpseudocode}
\usepackage{enumitem}

\usepackage{multirow}

\usepackage{booktabs}

\usepackage{tikz}
\usetikzlibrary{arrows.meta,positioning,calc}

\usepackage{filecontents}

\usepackage[nogroupskip,nonumberlist]{glossaries}
\setacronymstyle{long-short}
\makeglossaries
\newacronym{ais}{AIS}{Automatic Identification System}
\newacronym{gnss}{GNSS}{Global Navigation Satellite Systems}
\newacronym{ins}{INS}{Integrated Navigation System}
\newacronym{rf}{RF}{radio frequency}
\newacronym{cema}{CEMA}{cyber and electromagnetic activities}
\newacronym{sdr}{SDR}{software-defined radio}
\newacronym{nmea}{NMEA}{National Marine Electronics Association}
\newacronym{vdr}{VDR}{Voyage Data Recorder}
\newacronym{edr}{EDR}{Event Data Recorder}
\newacronym{tpm}{TPM}{Trusted Platform Module}
\newacronym{tsa}{TSA}{Time Stamp Authority}
\newacronym{pandi}{P\&I}{Protection and Indemnity}
\newacronym{ecdis}{ECDIS}{Electronic Chart Display and Information System}
\newacronym{ep}{EP}{Estimated Position}
\newacronym{pnt}{PNT}{Position, Navigation, and Timing}
\newacronym{ibs}{IBS}{Integrated Bridge System}
\newacronym{vhf}{VHF}{Very High Frequency}
\newacronym{ppi}{PPI}{Plan Position Indicator}
\newacronym{pps}{PPS}{pulse-per-second}
\newacronym{gpsdo}{GPSDO}{GPS-disciplined oscillator}
\newacronym{api}{API}{Application Programming Interface}
\newacronym{utc}{UTC}{Coordinated Universal Time}
\newacronym{gps}{GPS}{Global Positioning System}
\newacronym{pvt}{PVT}{Position, Velocity and Time}
\newacronym{ilu}{ILU}{Institute of London Underwriters}
\newacronym{iua}{IUA}{International Underwriting Association of London}
\newacronym{lma}{LMA}{Lloyd’s Market Association}
\newacronym{itc-hulls}{ITC-Hulls}{Institute Time Clauses}
\newacronym{iwsc}{IWSC}{Institute War and Strikes Clauses}
\newacronym{cl}{CL}{Market Clause}
\newacronym{ssot}{SSoT}{Single Source of Truth}
\newacronym{pcr}{PCR}{Platform Configuration Registers}
\newacronym{tcxo}{TCXO}{Temperature Compensated Crystal Oscillator}
\newacronym{ppm}{ppm}{parts per million}
\newacronym{tst}{TST}{Time-Stamp Token}
\newacronym{cdf}{CDF}{Cumulative Distribution Function}

\glsunset{api}
\glsunset{utc}
\glsunset{gps}
\glsunset{edr}

\newcommand\TOOLNAME[0]{RFbox}

\begin{document}

\date{}

\title{\Large \bf Spectrum of Liability: Forensically Sound Acquisition of Wireless Evidence for the Reconstruction of Maritime Cyber-Physical Incidents}



\ifdefined\anonymous
\else
\author{
{\rm Giacomo Longo}\\
CASD\\
School of Advanced Defense Studies, Italy\\
giacomo.longo@unicasd.it
\and
{\rm Alessandro Cantelli-Forti}\\
RaSS National Laboratory\\
CNIT, Italy\\
alessandro.cantelli.forti@cnit.it
\and
{\rm Alessio Merlo}\\
CASD\\
School of Advanced Defense Studies, Italy\\
alessio.merlo@unicasd.it
\and
{\rm Enrico Russo}\\
DIBRIS\\
School of Advanced Defense Studies, Italy\\
enrico.russo@unicasd.it} 
\fi

\maketitle

\begin{abstract}
Modern maritime navigation depends on mandatory wireless inputs, most notably \acrshort{gnss} and \acrshort{ais}, whose reception can be spoofed or jammed. When groundings or collisions occur, liability and safety investigations often turn on whether \acrshort{rf} interference corrupted bridge instruments, yet the artifacts typically available (\acrshort{vdr} outputs, \acrshort{ibs} logs, screenshots, decoded messages) are downstream, lossy, and easy to contest in adversarial proceedings. We introduce a forensically sound acquisition methodology that preserves upstream wireless evidence together with the minimal contemporaneous context needed to interpret it. Our system records coherent, multichannel IQ across relevant bands while simultaneously collecting time-aligned receiver and instrument-state metadata (e.g., decoded \acrshort{ais}/\acrshort{gnss} outputs and \acrshort{nmea} sentences), enabling later re-decoding and physical-layer reanalysis tied to what the onboard stack indicated at the time. To make the record independently auditable, we package samples and metadata into a single append-only forensic envelope that cryptographically binds per-stream hash chains with trusted-time anchors and periodic checkpoints, allowing third parties to detect tampering (modification, truncation, splicing, cross-case substitution) and to verify the claimed recording window. A prototype implementation shows the approach is practical: integrity mechanisms add negligible storage overhead and modest compute cost at operationally relevant multiband capture rates.
\end{abstract}

\section{Introduction}
\label{sec:intro}

Modern commercial shipping relies on radio-based navigation and situational-awareness systems whose outputs are treated, in practice and regulation, as authoritative inputs to safety-critical decisions. 
Two of the most consequential are \gls{gnss} (for position/time) and \gls{ais} (for cooperative traffic awareness). 
Yet both were designed without \textsl{adversarial \gls{rf} environments} in mind: \gls{gnss} reception is inherently fragile due to extremely low received power, and \gls{ais} messages are easy to manipulate at the protocol level \cite{balduzzi2014ais}.

This fragility is no longer hypothetical. 
Large-scale \gls{gnss} interference regions are continuously reported~\cite{gpsjam}, and recent casualties have been publicly attributed to suspected spoofing/jamming.
For instance, the containership MSC ANTONIA grounded near Jeddah in May 2025, with multiple independent analyses pointing to \gls{gnss} interference as a contributing factor~\cite{marinetraffic2025mscantonia}.
Similarly, the June 2025 collision between the tankers Front Eagle and ADALYNN occurred in a context of heightened concern about \gls{gnss} disruption in the Gulf/Hormuz area~\cite{nytimes2025hormuz}.

Beyond safety outcomes, insurers explicitly treat \gls{gnss} interference as an emerging operational and claims-handling issue, especially in geopolitical conflict zones~\cite{gard_gps_interference_conflict_zones}.
Yet, a crucial question quickly emerges: \textit{how can a shipowner (or an investigator) demonstrate, credibly and reproducibly, that a wireless attack contributed to an incident that occurred}? 
Today, the evidence typically available is downstream: operational logs, screenshots, and decoded message outputs.
These artifacts are often insufficient when the dispute depends on signal-level phenomena that are lost after demodulation and decoding (e.g., interference characteristics, timing consistency, and signal evolution over time). 
Moreover, because they are downstream and curated, they rarely provide an independently verifiable guarantee that the full \gls{rf} observation window was preserved without gaps or later alteration.

This paper addresses the resulting evidentiary gap. 
We present a forensic acquisition methodology that preserves upstream \gls{rf} evidence together with the minimal contemporaneous context needed to interpret it, and packages both into a single tamper-evident record that third parties can independently verify. 
The goal is not to attribute attackers or interpret legal causation, but to enable reproducible, adversarial scrutiny of ``what was received'' and ``what the onboard stack output at that time,'' without relying on one party’s subjective interpretation.


We provide the following contributions:
\begin{itemize}
    \item A method for forensically sound acquisition of wireless evidence that combines $(i)$ coherent multichannel \gls{rf} (IQ) recording to preserve the raw ether for re-decoding and physical-layer measurements and $(ii)$ in-band, time-aligned metadata (e.g., receiver state, \gls{nmea} sentences, and decoded \gls{ais}/\gls{gnss} artifacts) to reconstruct what the instruments actually observed.
    \item A verification mechanism for the forensic record that enables automatic checking of $(i)$ the integrity of the evidentiary record and its custody trail, and $(ii)$ the claimed time span of the recording, by cryptographically binding samples and metadata in a \gls{ssot} envelope with trusted time anchors.
    \item An extensive experimental evaluation quantifying storage and runtime overhead across representative multi-band acquisition profiles, showing feasibility at operationally realistic rates.
\end{itemize}

Although this paper is motivated by maritime \gls{gnss}/\gls{ais} interference, our core contribution is a protocol-level methodology for producing verifiable \gls{rf} evidence, applicable to other safety-critical wireless ecosystems such as aviation, intelligent transportation, and industrial communications (\S\ref{sec:generalized}).
\par\noindent\textbf{Structure of the paper.}
We first provide background (\S\ref{sec:background}) and a threat model (\S\ref{sec:threatmodel}), then motivate the evidentiary gap and define our goals  (\S\ref{sec:motivations}).
In \S\ref{sec:related}, we review related work.
We then present our methodology (\S\ref{sec:methodology}) and evaluate it in \S\ref{sec:evaluation}.
We conclude with a discussion (\S\ref{sec:discussion}) and closing remarks (\S\ref{sec:conclusion}).

\section{Background}
\label{sec:background}

\subsection{Maritime Bridge Systems and Data Flows}
\label{sec:maritimebridge}

Modern ship bridges are commonly organized as an \gls{ibs}~\cite{IMO_MSC_64_67,SOLAS_ChV_Reg19}, where multiple sensors and radio receivers provide navigation-relevant \emph{upstream} inputs that are processed, collected, and presented as \emph{downstream} outputs on dedicated consumers.
Among its integrated functions, an \gls{ibs} provides onboard systems with a \gls{pnt} function as a common navigation state and time base.

Core \gls{rf}[-based] inputs include: $(i)$ \gls{gnss} receivers, which provide World-wide \gls{pvt}~\cite{imo_msc74_69_gps_glonass,imo_msc401_95_multisystem_rnav}; 
$(ii)$ maritime Radar~\cite{IMO_MSC_192_79}, which provides local non-cooperative sensing of surrounding targets and environmental returns; and $(iii)$ \gls{ais}~\cite{ais}, a \gls{vhf} communication system with periodic cooperative transmissions of structured messages; for vessels, these include position, course, and speed.
In addition, bridges use inputs from conventional navigation sensors, including heading (gyrocompass), depth, and speed measurements.

Within the \gls{ibs}, derived outputs from sensors and radio receivers, together with equipment status, are exchanged over standardized shipboard data interfaces, commonly exposed as \gls{nmea}-style messages and networks (IEC 61162 family)~\cite{iec61162}. 
These interfaces feed different consumers, primarily bridge displays and recording units, which aggregate and transform such inputs into operator-facing products and logs.

On the display side, Radar information is commonly rendered as a \gls{ppi}, i.e., a processed visualization of radar returns, optionally augmented with tracked targets~\cite{imo_msc192_79_radar}. 
Radar installations may also support geo-referenced functions (e.g., stabilized presentation and chart overlays) that depend on \gls{pnt} and heading inputs. 
Alongside radar, \gls{ecdis} is the primary charting/navigation display~\cite{imo_msc530_106_ecdis}.
When the \gls{gnss} is unavailable or degraded, the above system may indicate an \gls{ep} by using dead reckoning from a previous fix and relying on onboard navigation sensors (e.g., heading and speed).
Finally, ships may carry a \gls{vdr}, which records selected bridge data items, including audio and replicas of key displays~\cite{imo_msc333_90_vdr}.

\subsection{Software Defined Radios}
\label{sec:sdr}
\Glspl{sdr} expose the received waveform as a stream of complex baseband samples.
For a channel tuned to center frequency $f_{LO}$, the analog \gls{rf} signal is mixed to baseband, low-pass filtered, and sampled at rate $f_s$, producing a discrete-time sequence
$$
x\bigl[k\bigr]=I\bigl[k\bigr]+jQ\bigl[k\bigr], k \in \mathbb{N}
$$
Representation of $x\bigl[k\bigr]$ depends on the hardware, but typical formats include signed 8-bit or 16-bit interleaved IQ (commonly referred to as \texttt{SC8} and \texttt{SC16}) and 32-bit float IQ (\texttt{CF32}).

Most \gls{sdr} platforms maintain a hardware time counter $T_{\mathrm{dev}}$ driven by the sampling clock. When the radio is locked to an external reference (typically a $10$ MHz square wave) and a $1$ Hz \gls{pps} signal, often provided by a \gls{gpsdo}, $T_{\mathrm{dev}}$ can be aligned to \gls{utc} and kept stable over long recordings; a common mode \emph{arms} the receiver and starts streaming on the next \gls{pps} edge\cite{uhd_sync}.
Letting $t_0$ denote the \gls{utc} time of that edge, the $k$-th complex sample is associated with $t(k)=t_0+\frac{k}{f_s}$, whereas without an external reference, $t_0$ is only relative to the device-local clock.

Although $x[k]$ is conceptually continuous, \gls{sdr} drivers deliver samples in variable-sized buffers;
typically accompanied by per-block metadata (e.g., $T_{\mathrm{dev}}$ for the first sample and status flags such as overflow/dropped-sample indications).
\Glspl{sdr} are operated through vendor drivers that expose \glspl{api} to configure \gls{rf}/analog parameters such as center frequency $f_{LO}$, sampling rate $f_s$, gain stages, filter bandwidth, antenna port, and to control the streaming lifecycle (start/stop, timed starts, and time-counter reads).

\section{Threat Model}
\label{sec:threatmodel}

We target modern vessels equipped with an \gls{ibs} and two \gls{rf} inputs (\gls{gnss} and \gls{ais}). 
Beyond standard bridge systems, the vessel hosts a dedicated forensic appliance implementing our acquisition methodology (\S\ref{sec:acquisition}), which we refer to as \emph{\TOOLNAME{}}.

\TOOLNAME{} includes a \gls{tpm} 2.0 acting as the hardware root of trust. 
A \emph{Reference Manifest} is created at commissioning and cryptographically bound to the provisioning authority (e.g., via a signature); it encodes the golden \gls{pcr} measurements~\cite{lyle2010trusted} for hardware topology, firmware configuration, and the authorized software stack. \TOOLNAME{} generates the local private key inside the \gls{tpm} and seals it to the expected PCR state; the \gls{tpm} releases the key only when current measurements match the Reference Manifest. 
We assume the \gls{tpm}’s security properties hold against software-only adversaries (e.g., Common Criteria EAL4+ or higher); invasive physical tampering is out of scope. 

We assume access to a \gls{tsa} that is not controlled by the adversary, and we rely on standard cryptographic assumptions (collision-resistant hashes and unforgeable signatures). 
Finally, we assume \TOOLNAME{} remains physically available after an incident (not stolen, hidden, or destroyed).

In this context, we distinguish three classes of actors, characterized by their \emph{incentives}, \emph{access} to relevant systems and evidence, and \emph{capabilities}.
\\
\par\noindent\textbf{Attackers.}
Attackers are external \gls{rf} adversaries whose actions target \gls{gnss} and/or \gls{ais} reception via \gls{cema}.
\emph{Incentives}: degrade situational awareness and induce unsafe behavior while preserving plausible deniability.
\emph{Access}: external-only, via \gls{rf} transmissions; no assumed access to onboard networks, bridge systems, or \TOOLNAME{}.
\emph{Capabilities}: can transmit interference and protocol-valid signals against \gls{gnss} and \gls{ais} receivers (e.g., jamming, spoofing, replay/meaconing), and can vary power, timing, and directionality to affect specific areas or time windows.

\par\noindent\textbf{Victims.}
Victims include \gls{rf}[-attack-affected] parties that rely on \gls{ibs} outputs or bear operational and financial exposure, such as the crew/operator and the shipowner.
\emph{Incentives}: minimize operational and legal exposure by supporting their account of events in the aftermath of an incident.
\emph{Access}: physical presence and operational control onboard; first-party custody of onboard artifacts immediately after an incident.
\emph{Capabilities}: can provide accounts of operations and decide what locally stored artifacts are produced post-incident; in adversarial settings, they may curate, omit, or modify such artifacts (e.g., truncation, selective preservation, or editing).

\par\noindent\textbf{Investigators.}
Investigators include authorities, insurers (and their claim assessors), and independent experts who review incidents in the context of proceedings or claims.
\emph{Incentives}: reach defensible conclusions and, where applicable, challenge evidentiary weaknesses or competing accounts in adversarial settings.
\emph{Access}: post-incident access to produced artifacts and analysis tooling; may obtain evidence through one or more parties rather than direct acquisition.
\emph{Capabilities}: can audit and reprocess provided data; in adversarial settings, can dispute completeness, integrity, and timing, and may perform post-processing on artifacts (e.g., truncation, re-ordering, splicing, or cross-case substitution) to support a preferred narrative.

\section{Motivations and Goals}
\label{sec:motivations}
In this section, we motivate the work with a representative case study, then show how the evidentiary gap plays out in safety investigations, claims handling, and formal proceedings. 
We conclude with concise goals and non-goals for our methodology.

\paragraph*{Motivating example.}
Gard is a specialised marine insurer providing \gls{pandi} liability cover~\cite{gard_shipowners_pi}.
The following case study is drawn from Gard’s published discussion of \gls{rf} interference in conflict zones~\cite{gard_gps_interference_conflict_zones} and is based on \gls{gnss}[-disruption] claims\footnote{referred to as ``\gls{gps}'' in the cited source} it has handled.

In \emph{Case Study~1}, a vessel lost \gls{gps} signal in the Red Sea and subsequently relied on the \gls{ecdis} \gls{ep} function (\textsection~\ref{sec:maritimebridge}) while navigating towards port.
Hours later, the crew observed unusual sea-surface coloration (mistakenly believed to be seaweed) and the vessel grounded on corals at 11.3~knots, causing substantial damage to the hull and the coral reef.
To establish the grounding location, the Flag State used external tracking data to confirm that the vessel was more than 40~NM away from the \gls{ecdis} \gls{ep}.
\\
\par\noindent\textbf{Safety investigation.}
Marine safety investigations aim to reconstruct causal factors through a methodical, evidence-based approach rather than adjudicating fault. 
This paradigm, as mandated by the IMO Casualty Investigation Code~\cite{imo_msc255_84}, requires a rigorous distinction between direct observations and post-hoc inferences. 
However, \gls{pnt}-related casualties, such as the grounding case study analyzed here, present a significant evidentiary challenge.

In this incident, the crew navigated using \gls{ecdis} \gls{ep} after a \gls{gnss} loss, with the final grounding location being established only later through third-party tracking. 
While investigators rely heavily on \gls{vdr} and \gls{ibs} logs to reconstruct the timeline, these artifacts are inherently downstream (\textsection~\ref{sec:maritimebridge}). 
Such records capture what was displayed and acknowledged on the bridge; yet, they fail to preserve the low-level signal conditions or receiver observables necessary to verify the root cause.
Consequently, when \gls{rf} interference is hypothesized as the trigger, the available evidence consists merely of indirect symptoms. 
Without independent access to the upstream \gls{rf} layer, it becomes nearly impossible to empirically distinguish between external interference and benign alternatives, such as receiver-specific faults or configuration issues.

\par\noindent\textbf{Insurance claims.}
The maritime insurance landscape is governed by model wordings produced by specialized industry bodies, primarily the \gls{ilu}, now \gls{iua}~\cite{iua_website}, and the \gls{lma}~\cite{lma_website}. 
These frameworks partition risks into distinct legal domains: the \gls{itc-hulls}~\cite{itc_hulls_1983} for standard marine perils, the \gls{iwsc}~\cite{iwsc_hulls_1983} for conflict-related hazards, and the more recent Cyber Attack Exclusion Clauses (CL380~\cite{cl380} and LMA5403~\cite{lma5403}).
Each framework is identified by a unique \gls{cl} reference number and serves as the deterministic logic for financial liability.

In our case study, the vessel lost reliable \gls{gnss} inputs, navigation degraded, and the vessel grounded. 
Even if this sequence is not disputed, the parties may still dispute the label of the cause.
The same grounding event can be argued as a Marine Peril (e.g., crew negligence under \gls{itc-hulls} cl. 6.2.3), a War Peril (e.g., hostile acts under \gls{iwsc} cl. 1.1), or a Cyber Event. 
In the latter case, cyber clauses are written broadly and depend on hard-to-answer questions, such as intent and whether the vessel was a target or a collateral victim.
As a result, bridge-level artifacts may be insufficient to determine how the loss should be framed under the relevant wordings, as they are downstream symptoms rather than evidence of the interference mechanism. 


\par\noindent\textbf{Formal proceedings and evidence handling.}
When disputes escalate into formal proceedings, the focus shifts from operational outputs to the evidentiary weight of the records. 
Latest industry guidelines~\cite{bimco2024guidelines} emphasize that rigorous evidence preservation is a critical requirement for casualty investigations and liability allocation.
To ensure completeness and reduce procedural challenges, such records must be handled through standardized forensic processes, such as ISO/IEC 27037~\cite{iso_iec_27037_2012}.
In our case study, moving from an observed \gls{gnss} loss to a legally defensible attribution of \gls{rf} interference depends on whether the underlying \gls{rf}[-layer] record is preserved under a robust chain of custody.
Accordingly, treating upstream \gls{rf} captures as forensic-grade digital evidence allows independent parties to assess integrity and the claimed recording window without relying on narrative interpretation.

\paragraph*{Goals.}
Accordingly, our methodology targets four goals designed to bridge the evidentiary gaps identified in the previous analysis:
\begin{enumerate}[label=\textbf{G$_\arabic*$}, leftmargin=*, align=left]
\item \textbf{Preserve upstream \gls{rf} evidence}. 
The system must capture coherent, multi-channel \gls{rf} I/Q data rather than relying solely on downstream logs that only record symptoms (e.g., the \gls{ecdis} \gls{ep} in Case Study 1). 
Such high-fidelity evidence enables the formal verification of \gls{rf} attack hypotheses, allowing investigators to re-analyze the physical layer even when bridge artifacts are ambiguous or their integrity is disputed.

\item \textbf{Bind \gls{rf} observations to onboard symptoms}. 
The system must capture the minimal downstream context (e.g., \gls{nmea} sentences and receiver settings) contemporaneous to the \gls{rf} capture. 
The contextual synchronization ensures the causal chain from external interference to the anomalies in onboard instrumentation remains auditable for investigators.

\item \textbf{Produce a verifiable single record}. 
The system must produce a \gls{ssot} bundle to provide an unbiased and complete evidence set for all stakeholders, satisfying the \emph{evidentiary weight} requirements of both formal legal proceedings and forensic investigations (e.g., ISO/IEC 27037).
This unified record ensures that integrity is independently checkable, preventing an interested party from performing an adversarial selection of logs (e.g., providing \gls{ecdis} data while omitting the \gls{rf} context).

\item \textbf{Provide verifiable time anchoring}. 
The system must anchor the \gls{ssot} bundle to a specific timeframe using cryptographic checkpoints. This is necessary to prevent post-hoc manipulation and prove the temporal alignment between \gls{rf} interference and system reactions (e.g., the 40~NM drift in Case Study 1), enabling objective liability assessment by third parties.



\end{enumerate}

\paragraph*{Non-goals.}
We do not attempt attribution or prove intent. We do not interpret insurance coverage, exclusions, or legal causation, nor do we seek to favor any involved actor.

\section{Related Work}
\label{sec:related}


Our work bridges the gap between forensic chain-of-custody, attestable logging, and \gls{rf} security. While prior research treats these domains in isolation, we provide an integrated, forensic-grade pipeline that captures upstream \gls{rf} observations alongside contemporaneous instrument context. This architecture ensures that both integrity and temporal claims are independently verifiable by third parties.

\paragraph{Forensic acquisition methodologies and evidence handling.}
Digital forensics has long emphasized disciplined acquisition and preservation processes to reduce procedural challenges and increase reproducibility, as codified in standards such as ISO/IEC~27037 and operational guides such as NIST~SP~800-86~\cite{iso27037,nist80086}. Recent work has also explored \emph{automatically verifiable} acquisition in domains beyond classic disk/memory imaging; for example, WEFT proposes a methodology for collecting web evidence with built-in verification primitives~\cite{cantelli_forti_weft2025}. In transportation, practitioner accounts of event/voyage data recorder (\gls{edr}/\gls{vdr}) forensics highlight that existing recorders are often $(i)$ downstream-oriented, $(ii)$ incomplete with respect to cyber-physical causality, and $(iii)$ not designed for adversarial disputes around tampering or selective disclosure~\cite{cantelli_forti_vdr2025}. These observations motivate moving evidence collection closer to the \emph{upstream} sources that determine what onboard systems perceive. Our methodology can be seen as a \gls{rf}-native counterpart: it preserves raw \gls{rf} (IQ) while binding it to the minimal context required to reconstruct what bridge systems observed, following a single ``forensic envelope'' abstraction rather than a collection of loosely coupled artifacts.

\paragraph{Tamper-evident logging and trusted auditing.}
A core building block is the extensive line of work on tamper-evident audit logging. Early foundations formalize secure audit logs under compromise and post-hoc scrutiny assumptions, typically using hash chaining and key-evolution to ensure tamper evidence~\cite{schneier1999secure}. Subsequent work improves efficiency and verifiability, e.g., via tree-based structures enabling compact inclusion/consistency proofs over large logs~\cite{crosby2009tamper}, and via append-only schemes designed to remain robust under partial compromise~\cite{yavuz2012appendonly}. Beyond cryptography-only approaches, systems research has leveraged trusted hardware to strengthen log integrity when the host OS is untrusted, including TEE-based log protection and practical trusted auditing frameworks~\cite{karande2017sgxlog,paccagnella2020custos}. More recent storage-centric and large-scale approaches (e.g., tamper-evident storage layers and shareable accountability proofs for IoT-scale logging) address retention and proof distribution~\cite{soriano2023sealfs,koisser2023accountability}. Our setting stresses these primitives in a distinct regime: high-rate, multi-stream \gls{rf} sample capture with periodic trusted-time anchoring, where $(a)$ the integrity object is not a low-rate event log but a sustained IQ stream, and $(b)$ verifiability must cover \emph{both} content integrity and claims about the recording window. Accordingly, we combine tamper-evident chaining with external time-stamping practices (e.g., RFC~3161 and related \gls{tsa} operational requirements) to support third-party verification of temporal bounds~\cite{rfc3161,etsi319421}.

\paragraph{Wireless deception and defense in navigation and situational awareness.}
A substantial body of security research demonstrates that \gls{rf} manipulation in navigation and awareness systems is practical and increasingly sophisticated. For \gls{gnss}, classical analyses characterize the conditions under which spoofing succeeds and thereby separate realistic threats from toy demonstrations~\cite{tippenhauer2011requirements}; surveys systematize the resulting attack/defense space~\cite{schmidt2016gnsssurvey}. Follow-on work shows that attackers can pursue \emph{plausibility} rather than obvious discontinuities (e.g., stealthy route-consistent manipulations), complicating post-incident interpretation from downstream logs alone~\cite{zeng2018gps}. Defensive techniques span single-receiver signal/constellation features~\cite{liu2021stars}, robust receiver designs and recovery under active spoofing~\cite{sathaye2022semperfi}, multi-sensor corroboration frameworks~\cite{dasgupta2022sensorfusion}, and opportunistic external signals used as side channels~\cite{oligeri2020iridium}; maritime-specific \gls{gnss} considerations further expose operational constraints of shipborne receivers~\cite{zalewski2014maritime}. For \gls{ais}, seminal evaluations document protocol-level weaknesses and practical attack surfaces~\cite{balduzzi2014ais}, while later work explores real-world spoofing case studies and systematic resilience testing of implementations~\cite{androjna2021aisspoof,khandker2022aisresilience}. In aviation, empirical studies on safety-critical \gls{rf} links (e.g., ILS and TCAS) underline that working attacks exist in realistic settings~\cite{sathaye2019ils,longo2024tcas}, and ``wireless witnessing'' proposes distributed, independent observers to detect and corroborate manipulation in ADS-B-based surveillance~\cite{jansen2021trustcrowd}; large-scale measurement studies similarly emphasize the gap between perceived and actual security in aviation \gls{rf} systems~\cite{strohmeier2016perception}. Collectively, these works motivate \emph{why} \gls{rf} evidence matters, but they are primarily oriented toward attack design, detection, or mitigation. Our contribution is complementary: we target \emph{post-incident reconstruction} and adversarial scrutiny by preserving the upstream \gls{rf} observations and binding them, verifiably, to the receiver/instrument state on which downstream systems acted  .

\paragraph{Maritime cybersecurity context.}
Finally, surveys and systematic reviews of maritime cybersecurity consistently identify \gls{gnss} and \gls{ais} as recurring high-impact dependencies within heterogeneous, legacy-heavy shipboard ecosystems~\cite{bolbot2022maritime,clavijo2024cybersecurityatsea}.
However, the literature largely treats evidence as an operational afterthought (e.g., \gls{vdr} outputs and bridge logs), leaving a gap in \emph{forensic readiness} when disputes hinge on what the receiver actually observed at the physical layer. Our work directly targets this gap by providing an evidence acquisition and verification pattern tailored to contested \gls{rf} environments.

\section{Methodology}
\label{sec:methodology}
\subsection{Assumptions}
\label{sec:assumptions}
\noindent\textbf{\TOOLNAME{ }clock initialization and drift.}
We assume that the \TOOLNAME{} clock is synchronized against a reliable reference in a trusted setting before each voyage (e.g., during harbor operations). 
After initialization, the clock operates in \emph{holdover mode} without further synchronization. 
Specifically, we assume the system is equipped with a Stratum 3~\cite{gr1244core} compliant high-precision \gls{tcxo} (e.g., SiTime SiT5356~\cite{sitime5356}) providing $\pm$0.1~\gls{ppm} stability.
This hardware choice limits the maximum theoretical drift to approximately 8.64~ms per day, resulting in a cumulative uncertainty of $\epsilon \approx 0.26$~s over a 30-day mission. 
Since $\epsilon$ remains well below the quantization threshold of standard filesystems (typically 1~s for POSIX-compliant file systems~\cite{carrier2005file}), the accumulated drift is unlikely to compromise the causal ordering of the evidence.


\noindent\textbf{Trusted time source.}
We assume access to an RFC3161-compliant \gls{tsa}\cite{rfc3161}.
We assume its published signing certificate (or certificate chain) is the intended trust root, and that the \gls{tsa} does not backdate tokens\cite{etsi_en319421_2025}.

To timestamp an object $x$, the requestor computes a digest
$
h \gets {\mathrm{alg}_{\mathrm{tsa}}}(x),
$
and submits a time-stamp request containing $(alg_{\mathrm{tsa}}, h)$.
The \gls{tsa} returns a time-stamp response that contains a signed \gls{tst} that cryptographically binds: $(i)$ the message imprint $(alg_{\mathrm{tsa}}, h)$, $(ii)$ a certified
generation time $\tau$, and $(iii)$ the \gls{tsa} identity, authenticated by the digital signature of the token and the \gls{tsa} signing certificate.

In the absence of external connectivity, we assume that \TOOLNAME{} can switch to an internal \gls{tsa}. 
This authority leverages the onboard clock (see above) to generate the tokens. 
To maintain forensic transparency, the internal \gls{tsa}  flags the \emph{holdover} state and populates the \emph{accuracy} via standard RFC3161 fields.
The latter represents the maximum temporal error $\epsilon$, calculated as the sum of the initial synchronization offset and the cumulative drift $(\pm 0.1 \text{ \gls{ppm}})$ since the last trusted harbor reference. 


\noindent\textbf{\gls{sdr} acquisition integrity.}
We assume the \gls{sdr} acquisition has an attested baseline, similar to a measurement instrument. 
Under this assumption, the hardware delivers a baseband stream consistent with its documented deterministic processing, while device identifiers, firmware, and driver versions are recorded as protected metadata to ensure auditability. 
Specifically, we assume each \gls{sdr} channel produces samples with a channel-local, monotonically increasing sample index.
Any non-deterministic discontinuities, such as overruns or drops, are reported as explicit metadata events. 
Finally, we assume the factory-specified clock drift appears in the attested baseline.

\subsection{Definitions}
\label{sec:definitions}
\paragraph{Configuration.}
An acquisition is specified by a configuration $\mathcal{C}$
\[
\mathcal{C} \coloneqq \langle N, N_{\mathrm{TSA}}, \{C_i\}_{i=1}^{N}, CS, CD, \mathrm{alg}_{\mathrm{blk}}, \mathrm{alg}_{\mathrm{tsa}} \rangle
\]
where:
(i) $N$ is the number of RF channels;
(ii) $N_{\mathrm{TSA}}$ is the checkpoint interval in seconds;
(iii) each channel specifier $C_i$
\[
C_i \coloneqq \langle \mathrm{idx}_i, \mathrm{sr}_i, \mathrm{sb}_i, \mathrm{lo}_i, \mathrm{drv}_i, \mathrm{cfg}_i, \mathrm{tpth}_i \rangle
\]
includes a channel index $\mathrm{idx}_i$, sample rate $\mathrm{sr}_i$, target block size $\mathrm{sb}_i$ samples, center frequency $\mathrm{lo}_i$, driver identifier $
\mathrm{drv}_i$, driver-specific settings $
\mathrm{cfg}_i$, and an optional temporary output directory $\mathrm{tpth}_i$;
(iv) $CS$ is the final case storage directory;
(v) $CD$ is arbitrary case description data;
(vi) $\mathrm{alg}_{\mathrm{blk}}$ is the hash algorithm used for blocks;
and (vii) $\mathrm{alg}_{\mathrm{tsa}}$ is a \gls{tsa}[-supported] digest.

\paragraph{Forensic envelope.}

\begin{figure*}
    \centering
    \includegraphics[width=\textwidth]{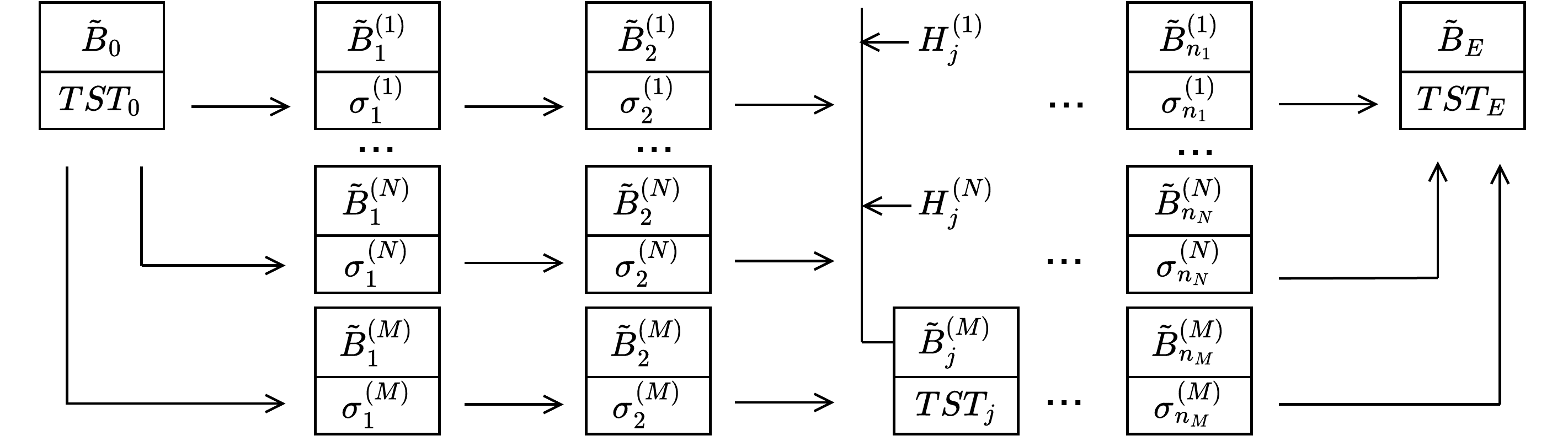}
    \caption{Structure of the forensic envelope.}
    \label{fig:envelope}
\end{figure*}

The forensic envelope $\mathcal{E}$, whose overall structure is depicted in \autoref{fig:envelope}, is the final artifact of the acquisition process:

$$
\mathcal{E} \coloneqq
\langle
  B_0,
  \{ \mathcal{S}_{C_i} \}_{i=1}^{N},
  \mathcal{S}_M,
  B_E
\rangle
$$

where $B_0$ is the initial block, $\mathcal{S}_{C_i}$ is the sequence of channel blocks $S_i$ for channel $i$, $\mathcal{S}_M$ is the metadata-block sequence, and $B_E$ is the end-of-recording block.

Each channel block sequence $S_c$ has the following structure
\begin{align*}
S_c \coloneqq \langle B^{(c)}_1, B^{(c)}_2, \ldots, B^{(c)}_{n_c}\rangle
\qquad c \in \{1,\ldots,N,M\}.
\end{align*}

Each persisted block $B$ is a tuple
\[
B \coloneqq \langle \mathrm{seq}, \mathrm{payload}, \sigma \rangle
\]
where $\mathrm{seq}\in\mathbb{N}$ is a strictly increasing sequence number within $S_c$ (or $-1$ for $B_0$ and $B_E$),
$\mathrm{payload}$ is a typed payload,
and $\sigma$ is the block signature.
For $B_0$ and $B_E$, $\sigma$ is a \gls{tsa} token.
For all intermediate blocks, $\sigma$ is a cryptographic hash linking the current block to the previous one.
Concretely, for each stream $c$, we maintain a running value  and update it as
$$
\sigma_i^{(c)} \coloneqq \; \mathrm{alg}_\mathrm{blk}(\sigma_{i-1}^{(c)} \; || \; \mathrm{enc}(\Tilde{B}_i^{(c)})\big)
$$
where $\Tilde{B}$ is the unsigned part of the block, and $\mathrm{enc}(\circ)$ is a canonical serialization.

To \emph{seed} each stream-specific chain, we bind it to the common start block and to the stream identifier:
$$
\sigma_0^{(c)} \coloneqq \mathrm{alg}_\mathrm{blk}(
  \mathrm{\tilde B_0} \; || \; \mathrm{enc(c)}
\big)
$$

\subsection{Acquisition}
\label{sec:acquisition}


We model acquisition as a protocol that, given an investigator-specified configuration $\mathcal{C}$, produces a tamper-evident forensic envelope $\mathcal{E}$ containing $(i)$ coherent multi-channel IQ recordings and $(ii)$ time-aligned contextual metadata, cryptographically sealed with trusted-time anchors.

\subsubsection{Architecture and data flows}

An acquisition instance consists of four logical components.

\begin{enumerate}
    \item \textbf{Coordinator.} Owns the global case state, emits global control events, and handles investigator commands.
    \item \textbf{Driver workers.} One per \gls{sdr} driver. Many \gls{sdr} stacks serialize device access or expose non-thread-safe driver APIs\cite{soapysdr_threading}; we therefore isolate each backend behind a single device-owner worker that performs all driver I/O and fan-out.
    \item \textbf{Channel workers.} One per \gls{rf} channel.
    Each worker receives the channel’s IQ stream and in-band events, packetizes them into blocks, applies hash chaining, and persists the resulting block sequence $S_{C_i}$ to a file under $\mathrm{SC}$ or at $\mathrm{tpth_c}$ if set.
    \item \textbf{Metadata worker.} Collects out-of-band context (e.g., \gls{nmea} sentences, decoded outputs, operator annotations), packetizes it into the metadata stream $S_M$, and stores trusted-time checkpoints.
    $S_M$ is persisted to a file under $\mathrm{SC}$ or at $\mathrm{tpth_M}$ if set.
\end{enumerate}

Communication is via message passing: driver workers emit tuples $\langle c, t_i, x\rangle$ where $c$ is the channel identifier, $t_i$ is a channel-local sample index (or device timestamp convertible to such an index), and $x$ is either an IQ segment or a metadata event.
This design keeps channels independent: packetization, hashing, and I/O can proceed in parallel across channels, and (when available) can be placed on distinct storage devices to reduce contention.

\subsubsection{Initialization and start sealing}

Acquisition starts by preparing the \gls{sdr} and attached devices for operation, crystallizing the acquisition context into the start block $B_0$, and time-stamping it.

\begin{enumerate}
    \item \textbf{Bring up and \gls{sdr} preparation.} The coordinator launches the sub-threads (driver/channel/metadata workers) and programs each \gls{sdr} front-end with its per-channel parameters ($\mathrm{drv}_i$ and $\mathrm{lo}_i$).
    \item \textbf{Freeze configuration and environment.} $B_0$ embeds the resolved configuration $\mathcal{C}$, relevant device identifiers (e.g., serial numbers), and acquisition-host attestation material (e.g., a \gls{tpm} quote) sufficient to later argue that a specific hardware/software stack produced the capture.
    
    $\tilde{B}_0$'s contents are $\langle \mathcal{C},\{{\mathrm{MD}_u\}_{u=1}^{U}}\rangle$, with every non-configuration information encoded within an arbitrary set of $U$ key-value entries $\mathrm{MD}_u=\langle k_u,v_u \rangle$.
    \item \textbf{Trusted-time anchor.} The coordinator obtains a \gls{tst} for $\Tilde{B}_0$ ($\mathrm{TST}_0$) which becomes the block signature of $B_0$.
    \item \textbf{Coordinated start.} After persisting $B_0$, the coordinator arms streaming and releases capture on the next synchronization boundary (e.g., a \gls{pps} edge when available, or an equivalent hardware-timed start).
    The metadata worker stores a \emph{start of recording} event $\mathrm{MD}_{\mathrm{sor}}$, tagged with the wall-clock time.
\end{enumerate}

Intuitively, $B_0$ commits to ``what was intended to be recorded'' (configuration) and ``with which environment'' (attestation), and binds these claims to an external time anchor before any samples are written.

\subsubsection{Sample streaming}

After the \emph{coordinated start}, acquisition enters the \emph{sample streaming} phase.

\paragraph{Channel workers.}
Each channel worker handles processing and persistence for a $S_x$.
It incrementally aggregates IQ samples and in-band metadata received from metadata threads until accumulating at least the configured target block size $\mathrm{sb}_c$, then emits a block.

Metadata events that arise within the \gls{sdr} pipeline such as gain changes, overflows, device timestamps, sample index gaps, or receiver status updates produced by the acquisition stack are represented as interval events $\langle \mathrm{MD}\footnote{$\mathrm{MD}$ is an arbitrary key-value pair, as defined above.}, t_{\mathrm{beg}}, t_{\mathrm{end}}\rangle$ in sample-index space.

Payload for a block in $S_x$ is thus $\bigl\langle \mathbf{x}, \{ \langle \mathrm{MD}, t_{\mathrm{beg}}, t_{\mathrm{end}}\rangle \} \bigr\rangle$, where $\mathbf{x}$ is a vector of IQ samples, followed by zero or more in-band metadata events.

\paragraph{Metadata worker.}
The metadata worker analogously emits to $S_M$ metadata blocks containing out-of-band events (e.g., \gls{nmea} sentences, operator notes, externally sourced WGS84 positions), each tagged with wall-clock time.
Payload for a block pertaining to $S_M$ consists of one or more arbitrary key-value pairs, i.e., payload for a metadata block containing $U$ entries is $\{{\mathrm{MD}_u\}_{u=1}^{U}}$.

\subsubsection{Periodic Trusted-Time Checkpointing}

Hash chaining detects modification, truncation, and splicing, but does not bind data to a creation time.
We therefore anchor the envelope to trusted time by embedding \glspl{tst} in $B_0$ and $B_E$ and, for long captures, checkpointing the multi-stream head state every $N_{\mathrm{TSA}}$ seconds during \emph{sample streaming}.

At checkpoint index $j$ (with $j$ increasing sequentially), the coordinator emits a checkpoint event that is picked up by all workers; upon receipt, each stream flushes any pending block(s), writes a snapshot request block with payload $\langle j \rangle$ to its stream, and reports its current head descriptor.

For stream $c$, the head is
\[
H^{(c)}_j = \langle c,\, \mathrm{seq}^{(c)}_j,\, \sigma^{(c)}_j \rangle,
\]
where $\mathrm{seq}^{(c)}_j$ denotes the latest block sequence number and $\sigma^{(c)}_j$ is the corresponding head digest; the metadata stream contributes an analogous head $H^{(M)}_j$. The coordinator then commits to the global state by forming the vector of heads
\[
H_j = \bigl[ H^{(1)}_j,\ldots,H^{(N)}_j, H^{(M)}_j \bigr]
\]
and computing an aggregate digest
\[
d_j \leftarrow \mathrm{alg}_{\mathrm{tsa}}\big(\mathrm{enc}(j \,\|\, H_j)\big).
\]
The digest $d_j$ is submitted to the \gls{tsa}, which returns $\mathrm{TST}_j$.
Finally, the coordinator persists the checkpoint by sending a record $\langle j, H_j, \mathrm{TST}_j\rangle$ to the metadata stream.

\subsubsection{Stop and final checkpoint}

Upon receiving a termination request, the coordinator broadcasts a control event to all workers and waits for an orderly drain.
Each channel worker and the metadata worker (i) flush any buffered samples/events, (ii) emit and persist any pending block(s), (iii) move the stream file from $\mathrm{tpth_c}$ to $\mathrm{CS}$ (if defined), and (iv) return their terminating head descriptor
$$
  H^{(c)}_{E} = \langle c,\; \mathrm{seq}^{(c)}_{E},\; \sigma^{(c)}_{E} \rangle \qquad c \in \{1,\ldots,N,M\}
$$
where $\mathrm{seq}^{(c)}_{E}=n_c$ is the last sequence number written for stream $c$ and $\sigma^{(c)}_{E}$ is the corresponding head digest.
The coordinator aggregates these heads into
$$
  H_{E} = \bigl[ H^{(1)}_{E},\ldots,H^{(N)}_{E},H^{(M)}_{E} \bigr]
$$
and constructs the end block $B_{E}$.
The unsigned payload $\tilde{B}_{E}$ commits to (i) $H_E$, i.e., a snapshot of the last blocks of all streams, and (ii) an on-disk manifest that binds stream identifiers to their corresponding files.
Finally, the coordinator obtains a \gls{tst} for $\tilde{B}_{E}$ ($\mathrm{TST}_E$) and uses it as signature for $B_{E}$.

\subsection{Verification}
\label{sec:verification}
Verification is a deterministic procedure that, for an envelope $\mathcal{E}$, asserts $(i)$ its unmodified status and $(ii)$ its temporal span.

\paragraph{Inputs and trust assumptions.}
Verification takes as input the case directory containing $\mathcal{E}$ and a \gls{tsa} trust root (certificate or chain) used to validate its signatures.
All cryptographic parameters (e.g., $\mathrm{alg}_{\mathrm{blk}}$ and $\mathrm{alg}_{\mathrm{tsa}}$) are read from the start block $B_0$.
Optionally, if the verifier is equipped to validate platform attestation material embedded in $B_0$ (e.g., \gls{tpm} quotes), it may additionally check that the acquisition host measurements match an expected policy.

\paragraph{Integrity verification checks.}

Validation of the integrity of $\mathcal{E}$ is performed by a verifier as follows:
\begin{enumerate}
\item \textbf{Start sealing.} Load $B_0$, deserialize its unsigned payload $\tilde B_0$, and validate the \gls{tst} against the \gls{tsa} trust root by checking that it certifies
\(\mathrm{alg}_{\mathrm{tsa}}(\tilde B_0)\).

\item \textbf{Per-stream well-formedness and contiguity.} For each stream $c\in\{1,\dots,N,M\}$, parse the block sequence $S_c$ and check that:
(i) every block deserializes; and
(ii) sequence numbers are contiguous and unique, i.e.,

$B^{(c)}_1.\mathrm{seq}=1$,
$\bigl\{ B^{(c)}_i.\mathrm{seq}=B^{(c)}_{i-1}.\mathrm{seq}+1 \;|\; \forall i\in\{2,\dots,n_c\} \bigr\}$,
and $B^{(c)}_{n_c}.\mathrm{seq}=n_c$ which implies no gaps and no duplicates.

\item \textbf{Per-stream hash-chain correctness.} For each stream $c$, recompute the stream seed
\(\hat\sigma^{(c)}_0 \gets \mathrm{alg}_{\mathrm{blk}}(B_0\ \|\ \mathrm{enc}(c))\)
and then, for each block $B^{(c)}_i\in S_c$, recompute the chained digest

$$
\hat\sigma^{(c)}_i \gets \mathrm{alg}_{\mathrm{blk}}\big(\hat\sigma^{(c)}_{i-1}\ \|\ \mathrm{enc}(\tilde B^{(c)}_i)\big),
$$
rejecting if any stored link $\sigma^{(c)}_i$ differs from $\hat\sigma^{(c)}_i$.

\item \textbf{End sealing and manifest consistency.} Load $B_E$, deserialize $\tilde B_E$, and validate its \gls{tst} by checking that it certifies
\(\mathrm{alg}_{\mathrm{tsa}}(\tilde B_E)\).
Then verify that $B_E$ commits to the locally computed terminal heads
\(\hat{H}^{(c)}_E=\langle c, n_c, \hat\sigma^{(c)}_{n_c}\rangle\) for all $c$.

\item \textbf{TSA checkpoint (snapshot) anchors.} Extract from the metadata stream the ordered list of checkpoint records
\(\mathcal{R}=\bigl[ R_1,\dots,R_k\bigr]\), where
\(R_j=\bigl[ \mathrm{id}_j, H_j, \mathrm{TST}_j\bigr]\)
and
\(H_j=\bigl[ H^{(1)}_j,\dots,H^{(N)}_j,H^{(M)}_j\bigr]\) with per-stream heads
\(H^{(c)}_j=\langle c,\mathrm{seq}^{(c)}_j,\sigma^{(c)}_j\rangle\).
For each $j\in\{1,\dots,k\}$, check:
\begin{enumerate}
\item \textbf{Sequentiality of snapshot identifiers.}
Check that checkpoint IDs are contiguous and strictly increasing.
$$
\min\mathrm{id} \in \{\emptyset\footnote{no snapshot is present.}, 1\} \wedge\ \forall j\in\{2,\dots,k\}:\ \mathrm{id}_j = \mathrm{id}_{j-1}+1,
$$
This rejects gaps, duplicates, and out-of-order snapshot identifiers.

\item \textbf{TSA binding.} Recompute the checkpoint digest
\(\hat d_j\gets \mathrm{alg}_{\mathrm{tsa}}(\mathrm{enc}(\mathrm{id}_j\ \|\ H_j))\)
and validate that $\mathrm{TST}_j$ certifies exactly $\hat d_j$ under the \gls{tsa} trust root.
Let $\tau_j$ be the $j$-th token generation time.

\item \textbf{Head consistency and monotonic progress.} For each stream $c$, check that (i) head positions are bounded and progress monotonically,
\(0 < \mathrm{seq}^{(c)}_1 < \cdots < \mathrm{seq}^{(c)}_k < n_c\), and (ii) the head digest matches the locally recomputed chain,
\(\sigma^{(c)}_j = \hat\sigma^{(c)}_{\mathrm{seq}^{(c)}_j}\).

\item \textbf{On-stream checkpoint markers.} For each stream $c$, verify that block $B^{(c)}_{\mathrm{seq}^{(c)}_j}$ contains a checkpoint marker for $\mathrm{id}_j$ (and that such markers are unique within the block), ensuring that the checkpoint is pinned to the stream order.

\item \textbf{Monotonicity of checkpoint times} Check that $\tau_0 \le \tau_1 \le \cdots \le \tau_k \le \tau_E$, where $\tau_0$ and $\tau_E$ are the generation times in $B_0$ and $B_E$.
\end{enumerate}
\end{enumerate}

\paragraph{Integrity guarantees.}
We say that a forensic envelope $\mathcal{E}$ is
\emph{integrity-verified} if it passes the integrity verification checks.
Under the cryptographic and trust assumptions in \S\ref{sec:assumptions}, acceptance
implies the following properties, each corresponding directly to one or more of the checks.

\begin{enumerate}
\item \textbf{Endpoint sealing is authentic (Checks~1 and~4).}
The start block $B_0$ and end block $B_E$ carry valid \gls{tsa} time-stamp tokens. Hence,
any post-acquisition modification of the committed acquisition context $\tilde B_0$
(e.g., configuration $\mathcal{C}$ and embedded environment/attestation material) or of the finalized envelope state $\tilde B_E$ (e.g., the committed terminal heads and
manifest) is detected by \gls{tsa} validation failure.

\item \textbf{Per-stream well-formedness and contiguity (Check~2).}
For every stream $c\in\{1,\dots,N,M\}$, the parsed block sequence
$S_c$ is structurally well-formed and its
sequence numbers are contiguous and unique.
Consequently, the presented stream cannot contain undetectable gaps or duplicates at the block level.

\item \textbf{Tamper-evident stream contents (Check~3).}
For every stream $c$, the verifier recomputes the seed and hash-chain links and requires exact agreement with the stored links.
Therefore, any post-hoc modification of a block payload, deletion/insertion of blocks, or reordering/splicing within a stream changes at least one recomputed link and is rejected (except with negligible probability under the assumed hash security).

\item \textbf{No cross-stream or cross-case cut-and-paste (Check~3).}
Because each stream chain is seeded with both the common start block $B_0$ and the stream identifier $c$, substituting a channel file with another channel’s file, or mixing blocks from a different case into the current case, breaks the recomputed chain and is
detected during verification.

\item \textbf{Consistent finalization and file binding (Check~4).}
The verified end block $B_E$ must commit to the locally computed terminal heads
$H^{(c)}_E$ for all streams.
Hence, the verified case directory is internally consistent: the stream files, their identifiers, and their terminal states match
exactly what was finalized at acquisition stop.

\item \textbf{Checkpoint record integrity (Check~5, if present).}
If periodic checkpoint records exist in $S_M$, then their
IDs are sequential, each \gls{tsa} token binds exactly to the advertised vector of heads, and each head matches the verifier’s recomputed chain at the advertised position; moreover, on-stream markers pin checkpoints to the stream order.
Thus, attempts to delete, insert,
duplicate, reorder, or alter checkpoint records are detected.
\end{enumerate}

Collectively, these properties mean that an integrity-verified envelope provides a single, internally consistent, tamper-evident source of truth for the recorded \gls{rf} samples and associated metadata, with any post-acquisition modification detectable by an independent verifier.

\paragraph{Temporal span guarantees.}
If verification succeeds, the verifier obtains a sequence of externally vouched times.
\(
\tau_0 \le \tau_1 \le \cdots \le \tau_k \le \tau_E
\)
Each checkpoint time \(\tau_j\) is tied to a \emph{validated multi-stream prefix} expressed directly in
block sequence numbers.
Let \(\mathbb{I} \coloneqq \{1,\ldots,N,M\}\) index the \(N\) \gls{rf} streams and the metadata stream, and let $\mathbf{s}_j$ be the \emph{head-position vector} for a given TSA snapshot record $R_j\in\mathcal{R}$.
\[
\mathbf{s}_j \coloneqq \bigl[\mathrm{seq}_j^{(c)}\bigr]_{c\in\mathbb{I}}
\qquad
\mathbf{s}_0 \coloneqq \bigl[0,\ldots,0\bigr]
\qquad
\mathbf{s}_E \coloneqq \bigl[n_c\bigr]_{c\in\mathbb{I}}
\]
Because verification enforces contiguity and uniqueness of sequence numbers within each stream, we can view
\(\mathrm{seq}_j^{(c)}\) as selecting the unique terminal block of stream \(c\) at the checkpoint.
We then define the induced \emph{multi-stream prefix} $E_{\preceq \mathbf{s}_j}$ as the tuple of per-stream prefixes

\[
E_{\preceq \mathbf{s}_j}
\;\coloneqq\;
\Bigl\langle \bigl[B^{(c)}_1,\ldots,B^{(c)}_{\mathrm{seq}_j^{(c)}}\bigr] \Bigr\rangle_{c\in\mathbb{I}}
\]
By the per-stream hash-chain recomputation and the checkpoint head-consistency checks, the verifier has validated
\(
\sigma_j^{(c)} = \hat{\sigma}^{(c)}_{\mathrm{seq}_j^{(c)}}
\)
for all \(c\in\mathbb{I}\), and the on-stream marker check pins \(\mathrm{id}_j\) to the terminal block
\(B^{(c)}_{\mathrm{seq}_j^{(c)}}\) in each stream.

Moreover, the \gls{tsa} binding check validates that \(TST_j\) certifies the digest $d_j$ at generation time \(\tau_j\).
Under a standard no-backdating assumption for the \gls{tsa}, each \(\tau_j\) therefore externally vouches that the
\emph{concrete set of blocks whose sequence numbers are at most \(\mathbf{s}_j\)} (i.e., \(E_{\preceq \mathbf{s}_j}\))
existed no later than \(\tau_j\); analogously, \(\tau_0\) vouches \(B_0\) and \(\tau_E\) vouches the final envelope
state committed in \(B_E\).

This yields two practically useful statements.
\begin{enumerate}
    \item \textbf{Non-backdating beyond anchors.} The acquisition context committed in $B_0$ existed no later than $\tau_0$, and the fully finalized envelope state (including all stream terminations) existed no later than $\tau_E$.
    \item \textbf{Bounded backdating within the case.} For each checkpoint $\mathrm{id}j$, all samples and metadata covered by the committed heads $H_{j}$ existed no later than $\tau_{j}$.
    With checkpoints spaced every $N_{TSA}$ seconds, any attempt to shift the claimed occurrence time of an \gls{rf} event by more than the checkpoint granularity requires either modifying a checkpoint record or recomputing a committed head, both of which are detected by the above checks.
\end{enumerate}

\subsection{Playback}
\label{sec:playback}

Playback re-emits a \emph{integrity-verified} forensic envelope $\mathcal{E}$  into \gls{sdr} transmit hardware to reproduce the recorded \gls{rf} observation under hardware-in-the-loop conditions. Unlike offline reprocessing (which consumes the stored IQ directly), playback is intended to $(i)$ exercise real receivers and bridge equipment, and $(ii)$ preserve the original timing relationships across bands and between \gls{rf} and auxiliary telemetry.

\paragraph{Inputs and preconditions.} Playback takes as input the case directory containing $\mathcal{E}$ and an investigator-specified mapping $\mu: \bigl[ c \bigr]_{c\in\mathbb{I}} \rightarrow \bigl[ \hat c \bigr]_{\hat c\in\mathbb{I}}$ from recorded channels to \gls{sdr} transmit chains.
From $B_0$, the player reconstructs the resolved configuration $\mathcal{C}$, including per-channel $(\mathrm{lo}_i,\mathrm{sr}_i)$ and the driver settings needed to reproduce the sample format and front-end configuration.

\paragraph{Architecture.}
A \emph{coordinator} thread configures hardware and controls start/stop.
For each stream $c\in{1,\dots,N}$, a \emph{stream reader} sequentially reads blocks and pushes their sample payloads into a bounded queue.
One \emph{TX worker} is instantiated per \gls{sdr} driver backend; it pulls from the queues of the channels it owns and performs all driver I/O.
A separate \emph{metadata replayer} consumes $S_M$ and delivers auxiliary events (e.g., \gls{nmea} sentences, operator annotations, externally sourced positions) to investigator-selected sinks.

\paragraph{Deterministic stream reconstruction.}
Each channel $c$ is represented as an ordered, gap-free sequence of blocks, and the envelope records terminal heads in $B_E$.
Consequently, playback reconstructs a unique contiguous IQ sequence by concatenation,
$$
\mathbf{x}^{(c)} \coloneqq \mathbf{x}^{(c)}_1 \Vert \mathbf{x}^{(c)}_2 \Vert \cdots \Vert \mathbf{x}^{(c)}_{n_c}
$$
where $\mathbf{x}^{(c)}_i$ denotes the sample vector carried in the payload of block $B^{(c)}_i$.
In-band metadata entries stored alongside samples (e.g., gain changes, device-timestamp reports) are indexed in sample space and are replayed deterministically at their recorded sample indices.
When an in-band event indicates a known discontinuity in the \gls{sdr} sample counter (e.g., an overflow), the player preserves the original timeline by inserting a synthetic ``gap'' segment  of the corresponding duration before continuing with the next recorded samples; this keeps subsequent in-band events aligned in sample-index space under the \emph{monotonic sample indexing} assumption.

\paragraph{Cross-channel coherence.}
Let $\mathrm{sr}_c$ be the sample rate for channel $c$.
Playback schedules transmission using a common hardware-timed start at time $t_{\mathrm{start}}$.
Concretely, for each channel $c$ and sample index $k\ge 0$, the intended transmit time is
$$
t^{(c)}(k) \coloneqq t_{\mathrm{start}} + \frac{k}{\mathrm{sr}_c}
$$
When the \gls{sdr} provides a shared timebase across TX chains, using the same $t_{\mathrm{start}}$ for all $c$ preserves the relative phase relationships captured during acquisition.
\paragraph{Metadata replay.}
The metadata stream $S_M$ carries out-of-band events tagged with wall-clock time.
The metadata replayer schedules each event at the same relative offset from the beginning of the capture, i.e., an event with recorded wall-clock time $t_j$ is emitted at local time $t_{\mathrm{start}} + (t_j - t_{\mathrm{ref}})$, where $t_{\mathrm{ref}}$ is the wall-clock annotation recorded in the first metadata record emitted after $B_0$, i.e. $\mathrm{MD}_{\mathrm{sor}}$.
Under the \emph{stable local wall clock} assumption, events played back as such do preserve the temporal relationships they had during acquisition.

\section{Evaluation}
\label{sec:evaluation}
We quantify the resource overhead introduced by our forensically sound acquisition pipeline, relative to a minimal baseline that only persists raw samples.
Our goal is to isolate the cost of envelope construction (serialization, hashing, and metadata handling) under representative multi-band acquisition workloads.
We do not evaluate waveform fidelity or decoder output quality.
Under our assumptions, fidelity is a property of the \gls{sdr} hardware and its driver, and both the baseline and our system exercise the same capture and replay primitives.

All experiments run on a workstation equipped with an AMD Ryzen 7900 CPU.
Our prototype is implemented in Rust\cite{rust} and uses BLAKE3\cite{OConnor2021BLAKE3} as its $\text{alg}_{\text{block}}$, SHA-256 for $\text{alg}_{\text{TSA}}$, OpenSSL as \gls{tsa}, and FlatBuffers\cite{flatbuffers} for $\mathrm{enc}(\circ)$.

We construct representative acquisition presets that emulate full-band capture for a set of maritime-relevant protocols spanning narrowband \gls{vhf} (\gls{ais}) to wideband \gls{gnss} signals.
Each protocol preset is defined by a center frequency $f_c$ and a nominal occupied bandwidth $B$. 
We emulate a full-spectrum recording for that band by selecting a complex sample rate $f_s = (1+\gamma)\,B \; \text{with } \gamma = 0.2$, i.e., a $20\%$ guard band over $B$.

Then, we define \emph{profiles} as subsets of presets that are acquired concurrently (one logical channel per protocol). Table~\ref{tab:eval:workloads} summarizes both the per-protocol presets and the evaluated profiles.

\begin{table}[t]
    \centering
    \caption{Protocol presets and acquisition profiles used in the evaluation. Profiles are subsets recorded concurrently.}
    \label{tab:eval:workloads}
    \footnotesize
    \begin{tabular}{lrrr}
        \toprule
        \multicolumn{4}{l}{\textbf{Protocol presets (single-channel)}}\\
        \midrule
        Protocol & $f_c$ [MHz] & $B$ [MHz] & $f_s$ [MS/s]\\
        \midrule
        \texttt{AIS\_a}        & 161.975  & \cite{ais}~0.025 & 0.030 \\
        \texttt{AIS\_b}        & 162.025  & \cite{ais}~0.025  & 0.030 \\
        \texttt{VHF}           & 159.000  & \cite{itur-m1084-5-2012}~3.100  & 3.720 \\
        \texttt{GPS\_L1}        & 1575.420 & \cite{itu-r_m1787-5_2024}~2.046  & 2.455 \\
        \texttt{GPS\_L2}        & 1227.600 & \cite{itu-r_m1787-5_2024}~2.046  & 2.455 \\
        \texttt{GPS\_L5}        & 1176.450 & \cite{itu-r_m1787-5_2024}~20.460 & 24.552 \\
        \texttt{GALILEO\_E1}    & 1575.420 & \cite{avila-rodriguez_galileo_signal_plan_2011}~24.552 & 29.462 \\
        \texttt{GALILEO\_E5a}   & 1176.450 & \cite{avila-rodriguez_galileo_signal_plan_2011}~20.460 & 24.552 \\
        \texttt{GALILEO\_E5b}   & 1207.140 & \cite{avila-rodriguez_galileo_signal_plan_2011}~20.460 & 24.552 \\
        \texttt{GALILEO\_E6}    & 1278.750 & \cite{avila-rodriguez_galileo_signal_plan_2011}~40.920 & 49.104 \\
        \midrule
        \multicolumn{4}{l}{\textbf{Acquisition profiles (multi-channel)}}\\
        \midrule
        Profile & \#ch. & Composition & $\sum f_s$ [MS/s]\\
        \midrule
        \texttt{AIS}        & 2  & \texttt{AIS\_a}, \texttt{AIS\_b}                    & 0.060 \\
        \texttt{VHF}& 1  & \texttt{VHF}                                       & 3.720 \\
        \texttt{GPS}        & 3  & \texttt{GPS\_L1}, \texttt{GPS\_L2}, \texttt{GPS\_L5} & 29.462 \\
        \texttt{GALILEO}    & 4  & \texttt{E1}, \texttt{E5a}, \texttt{E5b}, \texttt{E6} & 127.670 \\
        \texttt{GNSS}       & 7  & \texttt{GPS} $\cup$ \texttt{GALILEO}                & 157.133 \\
        \texttt{ALL}        & 10 & all presets above                                   & 160.913 \\
        \bottomrule
    \end{tabular}
\end{table}

Each channel reads samples from a deterministic tone generator producing SC8 IQs at rate $f_s$.
To exclude I/O bottlenecks while still measuring output size, all writes happen to a memory-backed filesystem.

For each of the $6$ profiles in Table~\ref{tab:eval:workloads}, we record $60$\,s and repeat each experiment $25$ times.

\subsection{Experiments}
\label{sec:eval:experiments}



Three experiments were conducted for each profile, as follows.
\textbf{(\textbf{E}$_1$) Baseline (raw recording).} We append the generated samples to file(s) without any additional processing, approximating the minimal cost of persisting the sample stream. \textbf{(\textbf{E}$_2$) Forensic pipeline.} We run the full acquisition pipeline described in the paper. \textbf{(\textbf{E}$_3$) Verification (after \textbf{E}$_2$).} We run our verifier over the produced envelope.

For $\textbf{E}_1$ and $\textbf{E}_2$ we measure: $(i)$ CPU utilization, $(ii)$ RAM usage, and $(iii)$ disk space consumption.
For each resource $R$, we normalize the forensic pipeline by the baseline:
\begin{equation*}
    \rho_R \coloneqq \frac{R_{\text{forensic}}}{R_{\text{baseline}}}
\end{equation*}
For $\textbf{E}_3$ we measure CPU utilization, RAM usage, and the verification speed w.r.t. realtime.

\subsection{Results}

\begin{table}[h]
\centering
\footnotesize
\caption{Footprint of forensic acquisition and verification.}
\label{tab:eval:absolute}
\begin{tabular}{lrrrr}
\toprule
\multirow{2}{*}{\textbf{Profile}} &
\textbf{CPU} &
\textbf{Disk} &
\textbf{Acq.\ mem} &
\textbf{Ver.\ mem} \\
 & [\%/core] & [MB/s] & [MB] & [MB] \\
\midrule
\texttt{AIS}     & 0.20 & 0.12 & 5.41  & 0.83 \\
\texttt{VHF}     & 5.67 & 7.44  & 6.68  & 5.61 \\
\texttt{GPS}     & 32.89 & 58.92  & 11.3  & 13.8 \\
\texttt{GALILEO} & 106.35 & 255.34   & 14.5  & 18.8 \\
\texttt{GNSS}    & 145.33 & 314.27   & 21.5  & 29.6 \\
\texttt{ALL}     & 147.97 & 321.86   & 24.7  & 33.2 \\
\bottomrule
\end{tabular}
\end{table}

\autoref{tab:eval:absolute} displays for each profile its associated overall CPU usage, disk throughput, acquisition, and verification RAM usage.
In aggregate, the most demanding profiles consume the equivalent of approximately 1.5 CPU cores.
Disk throughput peaks at $\approx 322$\,MB/s, while memory stays below $\approx 25$\,MB during acquisition and below $\approx 40$\,MB during verification.
\begin{figure}[t]
    \centering
    \includegraphics[width=1\linewidth]{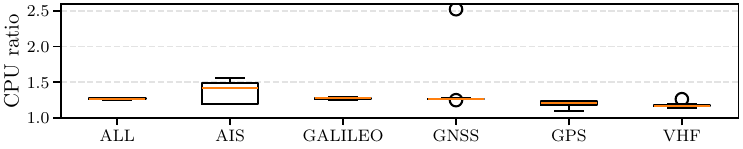}
    \caption{$\rho_R$ for CPU usage.}
    \label{fig:eval:cpu}
\end{figure}
\autoref{fig:eval:cpu} shows the CPU ratio w.r.t. baseline.
Across profiles, the overhead is typically around $1.2$--$1.3\times$, with the largest observed excursion reaching $\approx 2.5\times$ on the \texttt{GNSS} profile.
\begin{figure}[t]
    \centering
    \includegraphics[width=1\linewidth]{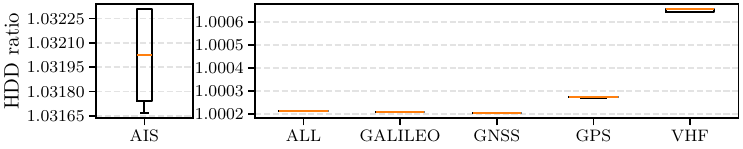}
    \caption{$\rho_R$ for storage usage.}
    \label{fig:eval:hdd}
\end{figure}
\autoref{fig:eval:hdd} shows the storage size, showing a negligible increase (at most $\approx3\%$) compared to the baseline.
\begin{figure}[t]
    \centering
    \includegraphics[width=1\linewidth]{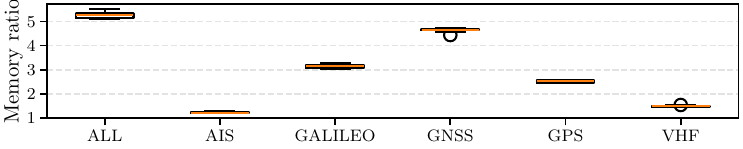}
    \caption{$\rho_R$ for memory usage.}
    \label{fig:eval:mem}
\end{figure}
\autoref{fig:eval:mem} depicts the memory usage ratio of the forensically sound pipeline w.r.t.\ the baseline, ranging from a near-baseline footprint in the \gls{ais} case ($\approx 1.2\times$) up to $\approx 5.3\times$ when all streams are enabled.
\begin{figure}[t]
    \centering
    \includegraphics[width=1\linewidth]{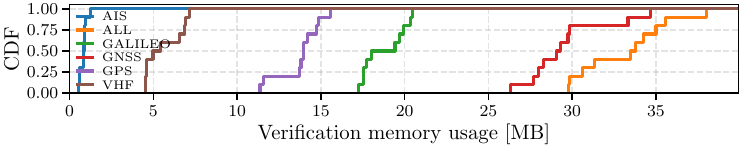}
    \caption{Memory usage CDF during verification.}
    \label{fig:eval:ver_mem}
\end{figure}
\autoref{fig:eval:ver_mem} shows the verifier memory usage, which stays below $\approx 40$\,MB (peaking at $\approx 38$\,MB) even when all streams are validated together.
\begin{figure*}[t]
    \centering
    \includegraphics[width=1\linewidth]{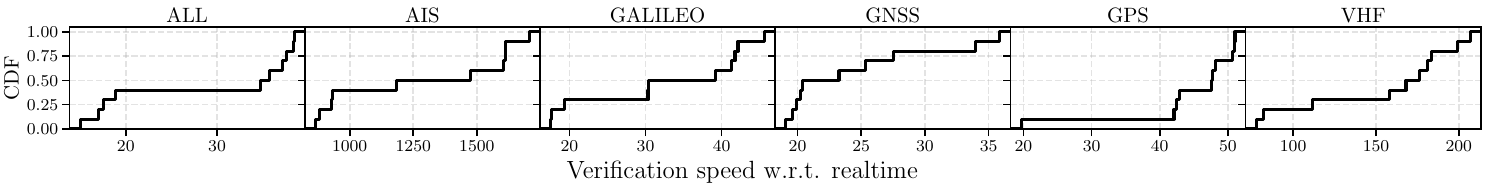}
    \caption{Verification speed w.r.t. realtime.}
    \label{fig:eval:ver_spd}
\end{figure*}
\autoref{fig:eval:ver_spd} shows the \gls{cdf} of the verifier speeds w.r.t. the length of the acquisition.
Verification ranges from $\approx 15\times$--$50\times$ faster than realtime on the wideband profiles up to $\approx 1000\!\times$ faster than realtime on \gls{ais}, slowing down roughly with the aggregate sample rate reported in \autoref{tab:eval:workloads}.



\section{Discussion}
\label{sec:discussion}

\paragraph{Performance considerations.}
Overall, the measurements indicate that making the capture \emph{verifiable} is inexpensive relative to the intrinsic cost of recording high-rate IQ streams.
Across all profiles, the forensic pipeline adds negligible storage overhead (worst case $\approx 3\%$) and significant yet bounded CPU overhead (at most $\approx 2.5\times$ baseline but less than 2 CPU cores overall), suggesting that serialization and hash chaining are not the throughput bottleneck.

The apparent memory inflation is best interpreted in absolute terms: while the memory \emph{ratio} can rise substantially, acquisition remains below $\approx 25$\,MB and verification stays below $\approx 40$\,MB, even when validating all streams together.
Verification is also practical for third parties: it runs from an order of magnitude to hundreds of times faster than real time, scaling primarily with aggregate sample volume.

Taken together, these results shift the deployment constraint away from cryptography and toward I/O: the heaviest profile sustains $\approx 322$\,MB/s of writes, so provisioning dedicated high-throughput storage (and avoiding contention with other logging tasks) is the dominant engineering requirement, not the integrity layer itself.

\paragraph{Trusted time anchoring: \gls{tsa} choice and the resulting guarantees.}

The envelope’s \emph{integrity} properties (tamper evidence, contiguity, cross-stream binding) follow from hash chaining and endpoint sealing.
However, \emph{temporal} guarantees, i.e., how strongly an external party can bound backdating, are shaped by the deployed \gls{tsa} model and by connectivity assumptions.


Binding periodic checkpoints to an external RFC3161-compliant \gls{tsa} yields the cleanest evidentiary story: timestamp tokens are verifiable under a public trust root and (under our assumptions) provide an external ``no-later-than'' bound for the committed envelope state at each checkpoint.
This makes it substantially harder for any single party to \emph{retroactively} shift the claimed time of \gls{rf} events without detection.

The operational cost is dependence on connectivity when requesting timestamp tokens. 
In practice, vessels may have intermittent satellite links, and adversaries capable of \gls{rf} interference may also impede the backhaul used to reach the \gls{tsa}.
When a request is delayed, the resulting token still certifies the envelope state, but only at the later token time; this weakens the backdating bound to ``no later than when connectivity resumed'', enlarging the window in which a malicious party could claim an earlier occurrence time.

Our local TSA service eliminates reliance on the Internet and provides \emph{internal} consistency: it attests to a time value coherent with the platform’s time base and signs checkpoint digests at high cadence.

However, this does \emph{not} provide an \emph{external} attestation of time. 
The trust model collapses to that of the shipboard acquisition/\gls{vdr} domain: a third party must trust that the local signer is correctly configured, uncompromised, and not backdating.
This posture is analogous to how shipboard recorders are typically treated in proceedings, i.e., the device self-attests its time base. 
Admissibility then depends on demonstrating correct configuration and absence of tampering, rather than on an independent time authority.


\paragraph{Continuous operation.}
Full-fidelity IQ acquisition is storage-intensive and, at first sight, appears costly and impossible to implement on commodity shipboard systems.
Even for multi-band, multi-channel settings, modern disks can sustain tens to hundreds of MB/s sequential writes, and commodity CPUs can hash at multi-GB/s rates using modern primitives, as reflected in our experimental results.
As such, such a system is within the realm of possibility.
These trends enable an operationally attractive mode: a \emph{certified ring buffer} of acquisitions.
The key observation is that the cryptographic overhead is small compared to sample volume, and the compute cost is dominated by streaming hashes and serialization, both of which map well to today’s CPUs.
The remaining practical constraint is I/O provisioning (dedicated disks, avoiding contention with other \gls{vdr} workloads), which is an engineering consideration rather than a fundamental barrier.

\paragraph{What we certify (and what we do not).}
The envelope provides strong guarantees against \emph{post-acquisition} tampering: an independent verifier can detect modification, truncation, reordering, cross-stream swapping, and (with TSA anchors) bound backdating at the checkpoint granularity.

This does \emph{not} imply that the recorded \gls{rf} signals were ``actually there'' in the physical world, nor that the \gls{sdr} stack was honest at acquisition time.
In other words, we certify that ``\emph{this is exactly what the recorder stored}'', not that ``\emph{this is exactly what the ether contained}''.
The distinction matters because an adversarial operator (or compromised acquisition host) could in principle:
$(i)$ inject synthetic IQ into the recording path,
$(ii)$ manipulate driver/firmware settings while reporting benign metadata,
or $(iii)$ record from an alternative antenna/front-end.

Our methodology already anticipates this by allowing attestation material to be embedded within $\mathrm{CD}$ within $B_0$, but meaningful \emph{authenticity} at acquisition time ultimately requires defense-in-depth:
secure boot for the host, signed \gls{sdr} firmware where available, measured driver stacks, physical tamper resistance, and (ideally) redundant sensing (e.g., independent receivers or dissimilar chipsets) to make fabrication detectable via inconsistency.
These measures strengthen the instrument model but do not eliminate the conceptual gap between ``recorded'' and ``physically observed''.

\paragraph{Integrating RF capture in \gls{vdr}s}
A central goal of the methodology is to bind raw \gls{rf} to the \emph{instrument state} that navigators and downstream systems acted upon (decoded outputs, \gls{nmea} sentences, receiver status, alarms).
In practice, these metadata feeds already traverse the ship’s sensor networks and are consumed by \glspl{vdr}.
This suggests a natural integration path: the metadata worker should be implemented as a first-class \gls{vdr} function (or a tightly coupled module), rather than an external add-on.

Embedding \gls{rf} acquisition into \glspl{vdr} has two advantages.
First, it reduces integration friction by reusing standardized sensor interfaces and existing retention policies.
Second, it aligns the physical custody story with existing regulatory practices for crash-protected recording systems.

The main engineering challenge is disk bandwidth: conventional \gls{vdr} inputs are orders of magnitude smaller than wideband IQ.
As such, designing a hypothetical \gls{vdr} with these features would be a complete redesign rather than the mere addition of another data stream.


\paragraph{Operational failure modes.}
High-rate acquisition inevitably faces overload conditions (I/O stalls, overruns).
The design choice to represent discontinuities as explicit in-band metadata is essential: it makes loss \emph{visible} and therefore debatable, rather than silently biasing reconstructions.
Still, in adversarial settings, dropped samples could be overzealously interpreted as the selective suppression of evidence. 
This can be mitigated by failing acquisition on such events.

\paragraph{Generalized formulation.}
\label{sec:generalized}
Although we focus on the maritime use case, the methodology itself is not maritime-specific: it is a protocol and domain-agnostic evidence pattern for any setting where $(i)$ safety/security outcomes depend on wireless signals, $(ii)$ disputes hinge on what the receiver actually observed (including physical-layer artifacts), and $(iii)$ third parties must be able to reproduce analyses under a verifiable chain of custody.

Examples include: aviation (\gls{gnss} interference\cite{tippenhauer2011requirements}, ADS-B/TCAS anomalies\cite{costin2012ghost,longo2024collision}, \gls{vhf} navigation/communications disputes\cite{strohmeier2016perception,sathaye2019wireless}), intelligent transportation (\gls{gnss} and V2X manipulation affecting automated or assisted driving\cite{zeng2018all,hasan2020securing,9868262}), industrial and critical infrastructure (wireless sensor/actuator networks, private LTE/5G links, microwave backhaul), and public-safety or enterprise environments (coverage, jamming, or spoofing allegations in incident response\cite{tucker2025detecting}).

\section{Conclusion}
\label{sec:conclusion}
We presented a forensically sound methodology to acquire maritime wireless evidence by capturing coherent multi-channel \gls{rf} alongside time-aligned instrument-state metadata, and packaging both into a single verifiable envelope. The envelope’s hash chaining and trusted-time anchoring make post-acquisition tampering and disputed timing claims independently checkable, improving dispute resilience in investigations and claims handling. Our prototype evaluation shows the approach is practical at operationally realistic rates, with modest overhead compared to raw recording. This provides a concrete path to reconstructing contested \gls{rf} incidents from upstream observations rather than relying on fragile downstream logs alone.

\clearpage
\section*{Ethical Considerations}
\label{app:ethics}

The acquisition workflow is \emph{passive} (receive-only): it does not transmit RF energy, probe remote systems, or otherwise interact with third-party infrastructure during recording.
Our evaluation does not rely on capturing over-the-air communications; instead, we use synthetic IQ streams generated locally.

\paragraph{Stakeholders.}
We consider the following stakeholders and impacts.
\emph{(i) Vessel operators/crew and shipowners (potential victims)} benefit from improved forensic readiness: the ability to produce a verifiable, independently auditable record can reduce wrongful liability and support safety investigations when RF interference is suspected.
\emph{(ii) Investigators, insurers, and courts} benefit from a reproducible evidence artifact that reduces disputes about integrity, completeness, and timing, and that discourages selective log curation.
\emph{(iii) Third parties whose transmissions may be incidentally captured} (e.g., nearby vessels, coastal stations, or other spectrum users) may face privacy and confidentiality risks if recordings include communications beyond what is necessary for incident analysis.
\emph{(iv) Adversaries} may be hindered by increased accountability and improved post-hoc scrutiny, but may also attempt to misuse the tooling for surveillance or to strengthen record-and-replay capabilities.
\emph{(v) The research team} is not exposed to human-subject risks in this study, as we do not collect personal data from human participants and do not run experiments on live third-party systems.

\paragraph{Potential harms and dual-use concerns.}
The primary ethical risk arises from the nature of raw RF capture: depending on front-end configuration (bandwidth, antenna placement, and channel selection), recordings can unintentionally include sensitive content (e.g., voice or other non-target transmissions), or enable downstream extraction of information not needed for the forensic question at hand.
A second risk is \emph{dual use}. While \gls{sdr} capture and replay are already well known, packaging and verification primitives could lower the operational friction for adversarial collection, sharing, and replay of \gls{rf} data.
Moreover, the multi-party nature of insurance/legal disputes can pressure parties into sharing more raw data than strictly necessary, potentially expanding exposure.

\paragraph{Mitigations.}
We mitigate these risks through responsible use constraints for the system.
\emph{Data minimization:} deployments should restrict acquisition to the narrowest frequency bands and bandwidths needed to answer the forensic question (e.g., \gls{ais} channels and relevant \gls{gnss} bands), avoiding broad-spectrum or opportunistic recording.
Configuration is explicit and auditable (as captured in the start block), enabling third parties to verify what was \emph{intended} to be recorded.
\emph{Evidence handling and access control:} raw \gls{rf} envelopes should be treated as sensitive forensic evidence.
Operators should store them under strict access controls, and follow established chain-of-custody procedures.
Sharing should be limited to authorized parties and governed by legal process and confidentiality agreements.
\emph{Selective disclosure limitations:} our integrity construction is optimized for whole-record verifiability; this can make ``reveal only the relevant minutes'' difficult without exposing additional context.
When partial disclosure is necessary, evidence should be shared via trusted escrow (e.g., court-appointed experts, as already customary in legal proceedings).
\emph{Playback safety:} although we describe playback for controlled reconstruction and hardware-in-the-loop testing, any transmission of recorded signals must be confined to shielded/cabled environments or otherwise compliant with radio regulations.
Playback should not be used over-the-air in operational bands.

\paragraph{Ethical principles and decision to publish.}
Under the Menlo principles: \emph{Beneficence} is served by improving safety investigations and reducing incentives/opportunities for evidence tampering; \emph{Respect for Persons} motivates minimizing incidental collection and enforcing strict confidentiality controls; \emph{Justice} is served by producing a single verifiable record that reduces asymmetric evidentiary power; and \emph{Respect for Law and Public Interest} requires that any real-world deployment comply with applicable spectrum and interception laws and established forensic practice.
We judge that the expected defensive and accountability benefits outweigh the residual risks, provided deployments follow the mitigations above and treat \gls{rf} captures as sensitive evidence rather than general-purpose telemetry.

\cleardoublepage
\bibliographystyle{plain}
\bibliography{refs}

\end{document}